\documentclass[runningheads]{cl2emult}

\usepackage{makeidx}  
\usepackage{graphicx} 
\usepackage{subeqnar} 
\usepackage{multicol} 
\usepackage{cropmark} 
\usepackage{eso}      
\makeindex            

\def\arcsec              {$^{\prime\prime}$} 
 
\newcommand \micron	{\mbox{$\mu m$~}}

\newcommand \etal   	{et al.}
\newcommand \degree  	{\mbox{$^\circ$}}
\newcommand \cent 	{$\omega$\thinspace Centauri}

\begin{document}
\title*{Tidal Tails around 20 Galactic Globular Clusters}
\toctitle{Tidal Tails around 20 Galactic Globular Clusters}
%
%
\titlerunning{Tidal Tails around 20 Galactic Globular Clusters}
%
\author{Georges Meylan\inst{1}
\and St\'ephane Leon\inst{2,3}
\and Fran\c{c}oise Combes\inst{3}
}
\authorrunning{G. Meylan et al.}
%
%
\institute{
	ESO, 
	Karl-Schwarzschild-Strasse 2, 
	D-85748 Garching, 
	Germany
\and 	ASIAA, 
	Institute of Astronomy and Astrophysics,
	Academia Sinica,\\
	P.O. Box 1-87,
	Nankang, Taipei 115, Taiwan
\and 	DEMIRM, 
	Observatoire de Paris, 
	61, Avenue de l'Observatoire,\\
	F-75015 Paris, France
}

\maketitle              


\section{Introduction}

In   addition to the effects  of   their internal dynamical evolution,
globular clusters suffer strong dynamical evolution from the potential
well of their host galaxy (Gnedin \& Ostriker 1997, Murali \& Weinberg
1997).   These   external  forces  speed   up the   internal dynamical
evolution  of these  stellar  systems, accelerating their destruction.
Shocks are caused by the tidal field  of the galaxy: interactions with
the disk, the bulge  and,  somehow, with  the giant molecular  clouds,
heat  up the outer  regions of each star   clusters.  The stars in the
halo are stripped  by the  tidal  field.   All  globular clusters  are
expected to have already  lost  an important  fraction of  their mass,
deposited in  the form of individual stars  in the  halo of the Galaxy
(see Meylan \& Heggie 1997 for a review).

Recent N-body simulations of globular clusters embedded in a realistic
galactic   potential  (Oh \& Lin    1992;  Johnston \etal\ 1999)  were
performed in order  to  study the amount  of  mass loss for  different
kinds of orbits  and  different  kinds  of clusters, along   with  the
dynamics and the mass   segregation in tidal tails.  Grillmair  \etal\
(1995) in an observational analysis of star counts  in the outer parts
of a few galactic globular  clusters found extra-cluster overdensities
that they associated partly with stars stripped into the Galaxy field.

We  present hereafter  our  study of the  2-D  structures of the tidal
tails associated with 20 galactic globular clusters, obtained by using
the wavelet  transform  to detect weak structures   at large scale and
filter the  strong  background  noise for  the   low galactic latitude
clusters   (Leon, Meylan \&  Combes  1999).   We  also  present N-body
simulations of globular clusters in orbits around the Galaxy, in order
to   study quantitatively  and   geometrically the  tidal effects they
encounter (Combes, Leon \& Meylan 1999).

\section{Observations and Data Reduction}

Our sample clusters share different   properties or locations in   the
Galaxy, with various masses and structural parameters. It is of course
necessary to have very wide  field imaging observations, consequently,
we obtained, during  the years 1996 and  1997, photographic films with
the ESO Schmidt telescope.  The field of view is of $5.5\degree \times
5.5\degree$ with  a scale of 67.5\arcsec/mm.   The  filters used, viz.
BG12 and RG630,  correspond to $B$ and  $R$, respectively.  All  these
photographic films were digitalized using the MAMA scanning machine of
the Observatoire de Paris, which provides a pixel  size of 10 \micron.
The  astrometric performances of the  machine  are described in Berger
\etal\ (1991).

The next  step -- identification all point  sources in these frames --
was performed  using SExtractor (Bertin \&   Arnouts 1996), a software
dedicated to  the  automatic analysis of  astronomical images  using a
multi-threshold   algorithm  allowing   good  object deblending.   The
detection  of the stars   was done at  a   3-$\sigma$ level above  the
background. This software, which can deal with huge amount of data (up
to 60,000 $\times$ 60,000 pixels) is not suited for very crowded field
like the centers of the  globular clusters, which were simply ignored.
We  performed a star/galaxy    separation   by using  the  method   of
star/galaxy magnitude vs. log(star/galaxy area).

For each  field,  we constructed a  $B$  vs.   $(B-V)$ color-magnitude
diagram, on   which  we performed   a  field/cluster  star  selection,
following the  method of Grillmair \etal\  (1995), since cluster stars
and field stars exhibit different colors.  In this way we can identify
present and past cluster  members from the  fore- and background field
stars by identifying in  the CMD the  area occupied by cluster  stars.
The envelope of this area is empirically chosen so  as to optimize the
ratio  of cluster  stars to  field  stars  in  the relatively sparsely
populated outer regions of each cluster.

\section{Wavelet Analysis}

With the assumption that  the data can be viewed  as a sum of  details
with different  typical scale  lengths,  the  next step  consists   of
disentangle  these details using  the space-scale analysis provided by
the Wavelet Transform (WT, cf.  Slezak \etal\ 1994; Resnikoff \& Wells
1998).  Any observational signal includes also some noise, which has a
short  scale length.  Consequently the noise  is higher  for the small
scale wavelet coefficients.   We performed  Monte-Carlo simulations to
estimate the noise at each scale and apply a 3-"$\sigma$" threshold on
the wavelet coefficients  to  keep only the  reliable structures.   In
this way it is possible to subtract the short-wavelength noise without
removing details from the signal which has longer wavelengths.

The remaining overdensities of the cluster-like stars, remaining after
the application of the   wavelength  transform analysis to  the   star
counts,  are  associated with the  stars  evaporated from the clusters
because of dynamical relaxation and/or tidal stripping by the galactic
gravitational field.

It is worth emphasizing that we have taken  into account the following
strong  observational biases: (i)   bias   due to  the  clustering  of
galactic  field stars; (ii) bias  due to the  clustering of background
galaxies; (iii) bias due to  the fluctuations of the dust  extinction,
as observed in the IRAS 100-\micron map.

\begin{figure}
\centering
\includegraphics[width=.8\textwidth]{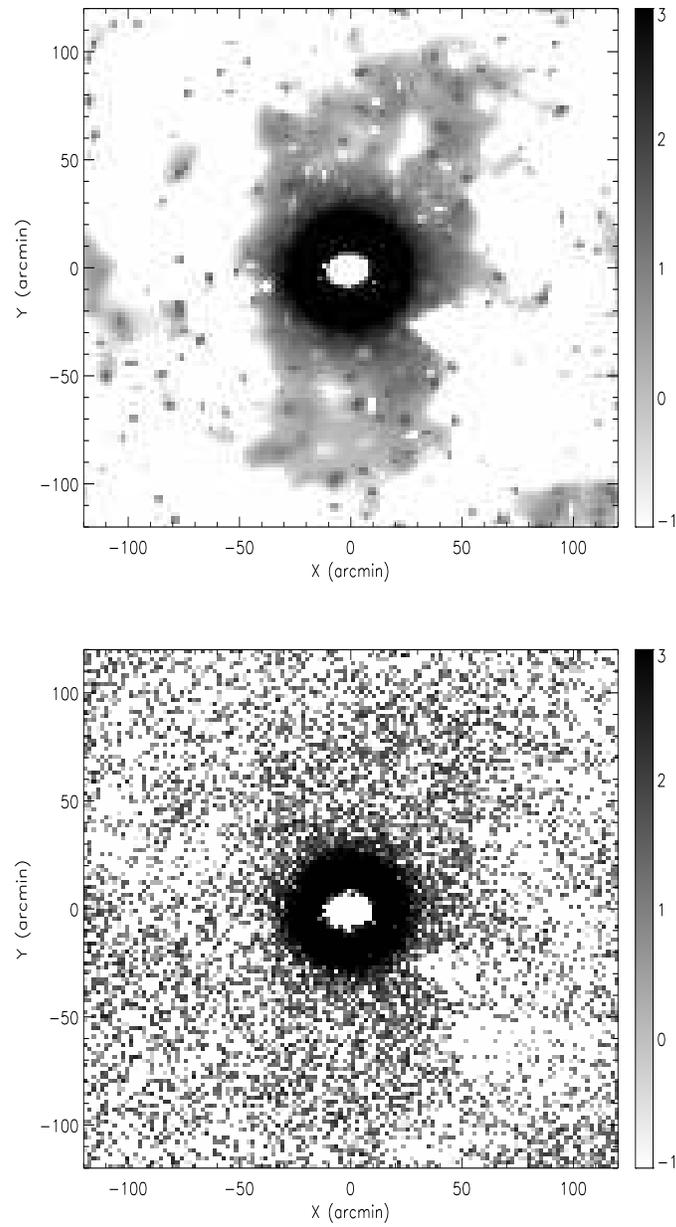}
\caption[]{NGC~5139  $\equiv$  \cent.  In   the upper  panel, filtered
image of  color-selected  star-count overdensities using  the  Wavelet
Transform to be compared with the  raw star counts in the lower panel.
The upper  panel displays the full  resolution using the whole  set of
wavelet planes. }
\label{}
\end{figure}

\section{Observational Results}

NGC~5139 $\equiv$ \cent, the   most massive galactic globular  cluster
(Meylan  \etal\ 1995), currently crossing the  disk plane, is a nearby
globular cluster located at a distance  of 5.0~kpc from  the sun.  Its
relative proximity  allows to  reach  the main sequence for  the  star
count  selection.  We   estimate, taking   into  account the  possible
presence of  mass segregation in its   outer parts, that about  0.6 to
1~\% of its mass has been lost during the current disk shocking event.
Although this cluster has  one of our  best tail/background S/N ratio,
it is by far not the only one exhibiting tidal tails.

Considering all   20 clusters of  our sample,  we  reach the following
conclusions  (see   Leon,   Meylan  \&  Combes 1999    for a  complete
description of this work):
\begin{itemize}

\item All   the clusters observed,  which do  not suffer   from strong
observational biases, present  tidal  tails, tracing  their  dynamical
evolution in the  Galaxy (evaporation, tidal shocking, tidal torquing,
and bulge shocking).

\item The clusters in the following sub-sample (viz. NGC~104, NGC~288,
\break  NGC~2298, NGC~5139, NGC~5904,  NGC~6535, and NGC~6809) exhibit
tidal extensions  resulting from a recent   shock, i.e.  tails aligned
with the tidal field gradient.

\item The  clusters  in another sub-sample  (viz.  NGC~1261, NGC~1851,
NGC~1904, NGC~5694,  NGC~5824, NGC~6205, NGC~7492, Pal~5,  and Pal~12)
present  extensions which are  only tracing  the  orbital path of  the
cluster with various degrees of mass loss.

\item NGC~7492 is a striking case because of  its very small extension
and  its high  destruction rate driven   by the galaxy as  computed by
Gnedin  \&  Ostriker (1997).   Its    dynamical ``twin''  for  such an
evolution, namely Pal~12, exhibits, on the contrary, a large extension
tracing its orbital path, with  a  possible shock which happened  more
than 350~Myr.

\item The presence of a break in  the outer surface density profile is
a reliable indicator of some recent gravitational shocks.
\end{itemize}

Our recent CCD observations with the Wide  Field Imager at the ESO/MPI
2.2-m  telescope will soon   provide improved results,  because of the
more accurate CCD   photometry.   These observations will  allow  more
precise observational estimates of  the mass loss rates for  different
regimes of galaxy-driven cluster evolution.

\section{Numerical Simulations}

We tried  with extensive numerical  simulations to reproduce the above
observations.  We  performed N-body simulations  of globular clusters,
in  orbits around  the Galaxy, in  order  to study quantitatively  and
geometrically the tidal effects they encounter.   The N-body code used
is an FFT algorithm,  using the method of   James (1977) to avoid  the
periodic images.  With N = 150,000 particles,  it required 2.7s of CPU
per time step on a Cray-C94.

\begin{figure}
\centering
\includegraphics[width=.84\textwidth]{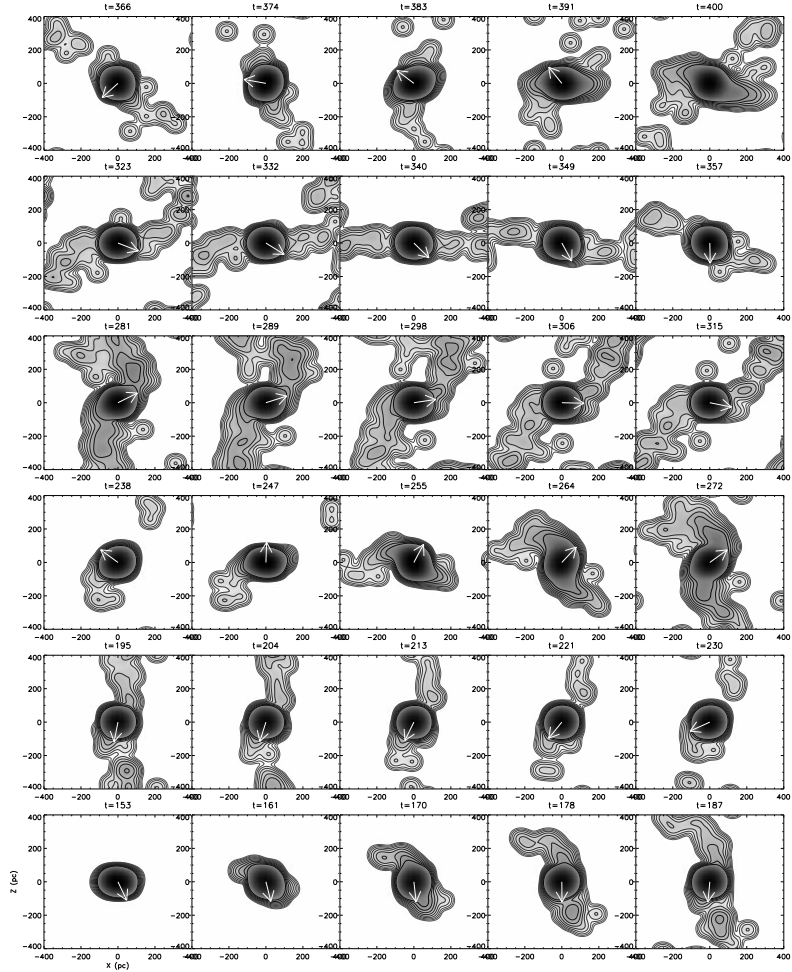}
\caption[]{  Tidal tails mapped at  different epochs  with the wavelet
algorithm  applied    to  one of   our   simulations.   The  direction
perpendicular  to the galactic  plane is indicated by  the arrow.  The
time sequence starts with the lower-left panel and ends with the upper
right one.  The third panel  exhibit tails which are quite reminiscent
of what is observed in NGC~5139 $\equiv$ \cent\ (see Fig.~1).}
\label{}
\end{figure}

The  globular   clusters are   represented by   multi-mass King-Michie
models, including mass segregation  at initial conditions.  The Galaxy
is modelled as  realistic as possible,  with  three components, bulge,
disk and dark halo: the bulge is a spherical  Plummer law, the disk is
a Miyamoto-Nagai model, and the dark matter halo is  added to obtain a
flat Galactic rotation curve.

The  main conclusions of  our simulations can  be summarized as follows
(see Combes, Leon  \& Meylan 1999  for a complete  description of this
work):
\begin{itemize}
\item All  runs show that the  clusters are always surrounded by tidal
tails  and debris.  This  is also true for  those that suffered only a
very slight mass loss.   These unbound particles distribute in volumic
density like a power-law as a function of  radius, with a slope around
--4. This slope  is much  steeper than  in the observations  where the
background-foreground contamination dominates at very large scale.
\item These tails are preferentially composed of low mass stars, since
they are coming  from the external radii of  the cluster; due  to mass
segregation built  up  by  two-body  relaxation,  the  external  radii
preferentially gather the low mass stars.
\item For sufficiently high  and rapid mass  loss, the cluster takes a
prolate shape, whose major axis precess around the z-axis.
\item When the tidal tail is very long (high mass loss) it follows the
cluster orbit: the  observation of the tail  geometry is thus a way to
deduce cluster   orbits.   Stars   are not  distributed  homogeneously
through the  tails, but form clumps,  and the densest of them, located
symmetrically  in  the  tails,   are  the  tracers of   the  strongest
gravitational shocks.
\end{itemize}
Finally, these   N-body   experiments help  to   understand the recent
observations of extended    tidal  tails  around  globular    clusters
(Grillmair et al. 1995, Leon et al. 1999): the systematic observations
of the geometry  of these tails should  bring much information  on the
orbit,  dynamic,  and mass loss  history of  the  clusters, and on the
Galactic structure as well.


\clearpage
\addcontentsline{toc}{section}{Index}
\flushbottom
\printindex

\end{document}